\documentclass[prb,twocolumn,10pt]{revtex4-1}

\usepackage{graphicx}
\usepackage{verbatim}
\usepackage{comment}
\usepackage{amssymb}
\usepackage{stmaryrd}

\begin{document}

\title{Thermodynamics of the Antiferromagnetic Heisenberg Model on the Checkerboard Lattice}
\author{Ehsan Khatami and Marcos Rigol} 
\affiliation{Department of Physics, Georgetown University, Washington DC, 20057, USA}

\begin{abstract}

Employing numerical linked-cluster expansions (NLCEs) along with exact diagonalizations 
of finite clusters with periodic boundary condition, we study the energy, specific heat, 
entropy, and various susceptibilities of the antiferromagnetic Heisenberg model on the 
checkerboard lattice. NLCEs, combined with extrapolation techniques, allow us to access 
temperatures much lower than those accessible to exact diagonalization and other series 
expansions. We show that the high-temperature peak in specific heat decreases as the 
frustration increases, consistent with the large amount of unquenched entropy in the 
region around maximum classical frustration, where the nearest-neighbor and next-nearest-neighbor 
exchange interactions ($J$ and $J'$, respectively) have the same strength, 
and with the formation of a second peak at lower temperatures. The staggered susceptibility 
shows a change of character when $J'$ increases beyond $0.75J$, implying the disappearance 
of the antiferromagnetic order at low temperatures. For $J'=4J$, in the limit 
of weakly coupled crossed chains, we find large susceptibilities for stripe and N\'{e}el 
order with ${\bf Q}=(\pi/2,\pi/2)$ at intermediate temperatures. Other magnetic and bond 
orderings, such as a plaquette valence-bond solid and a crossed-dimer order suggested by 
previous studies, are also investigated.

\end{abstract}

\maketitle

\section{Introduction}

The checkerboard lattice is a unique two-dimensional (2D) system of great current 
interest. The next-nearest-neighbor (NNN) interactions, which are present on every other 
plaquette in a checkerboard pattern, not only can impose frustration and drive the 
system to exotic ground states but also provide a great tool for numerical and 
analytical investigators to study the evolution of physical properties in transitions 
between different geometries. For instance, in the limit of weak NNN interactions, it 
is expected that the physics associated with the simple square lattice is dominant. In 
the antiferromagnetic Heisenberg (AFH) model, this means a tendency toward long-range 
N\'{e}el ordering at temperatures smaller than the characteristic energy scale set by the 
nearest-neighbor (NN) magnetic exchange interaction, $J$. Whereas a ferromagnetic (negative) 
NNN exchange interaction, $J'$, favors this N\'{e}el ordering, an antiferromagnetic 
(positive) one introduces frustration and, thus, new types of ordering such as a valence-bond 
solid emerge. In the fully frustrated region where $J\sim J'> 0$, the lattice is 
a projection of the three-dimensional corner-sharing tetrahedrons (pyrochlore lattice) 
onto a 2D lattice. The other interesting limit is $J'\gg J$, where the 2D lattice is 
practically reduced to weakly coupled crossed chains, and physical properties are dominated 
by those of the one-dimensional (1D) system. Moreover, by eliminating certain bonds, one can even 
turn the focus from the square basis of the underlying lattice to a triangular one that 
can capture the geometry of the Kagom\'{e} lattice.

The problem of the frustrated AFH model on the checkerboard lattice has its roots in
early studies on its three-dimensional counterpart, the pyrochlore lattice. The latter 
system was originally studied by Harris {\em et al.} \cite{A_harris_91} using quantum field 
theory. They ruled out the possibility of a phase 
with long-range spin correlations but found strong correlation between NN spins, 
suggesting a dimerized ground state. A few years later, using perturbative expansions 
and exact diagonalization, Canals and Lacroix~\cite{b_canals_98} concluded that the 
ground state is a spin-liquid with correlations that decay exponentially by distance. 
Around the same time, another study by Isoda and Mori,~\cite{M_Isoda_98} in which a 
bond-operator approach was used, suggested a resonant-valence-bond-like plaquette phase.

So far, the magnetic properties of the checkerboard lattice have been the focus of many 
theoretical studies,~\cite{W_Brenig_02,O_Starykh_02,W_Brenig_04,J_Fouet_03,O_Tchernyshyov_03,O_Starykh_05,
r_moessner_04,E_Berg_03,J_Bernier_04,P_Sindzingre_02,r_singh_98,b_canals_02,M_arlego_07,S_Moukouri_08,
S_Palmer_01,E_Lieb_99} with compelling evidence that the ground state for $J'=J$ (the planar pyrochlore) 
is a plaquette valence-bond solid (P-VBS) with long-range quadrumer 
order.~\cite{W_Brenig_02,W_Brenig_04,J_Fouet_03,O_Tchernyshyov_03,J_Bernier_04,O_Starykh_05,r_moessner_04}
This was shown by means of strong-coupling expansion,~\cite{W_Brenig_02,W_Brenig_04}
exact diagonalization,\cite{J_Fouet_03} as well as mean field theory \cite{r_moessner_04}
and a quadrumer boson approximation.~\cite{O_Starykh_05}

In the limit of $J'\ll J$, the existence of the long-range N\'{e}el order has also 
been established.\cite{P_Sindzingre_02,J_Fouet_03,r_singh_98,b_canals_02} Semiclassical 
approaches, such as the linear spin-wave,~\cite{r_singh_98,b_canals_02} and numerical 
results~\cite{P_Sindzingre_02,J_Fouet_03} predict the stability of antiferromagnetic (AF) 
long-range order for $J'/J\lesssim 0.75$. However, this number is different in other 
studies that associate the instability of the P-VBS phase, as $J'$ is reduced, with 
the transition to the N\'{e}el state ($5/8$ in Ref.~\onlinecite{O_Starykh_05}, and 
$0.88-0.94$ in Ref.~\onlinecite{W_Brenig_04}).

\begin{figure}[t]
\centerline {\includegraphics*[width=3.3in]{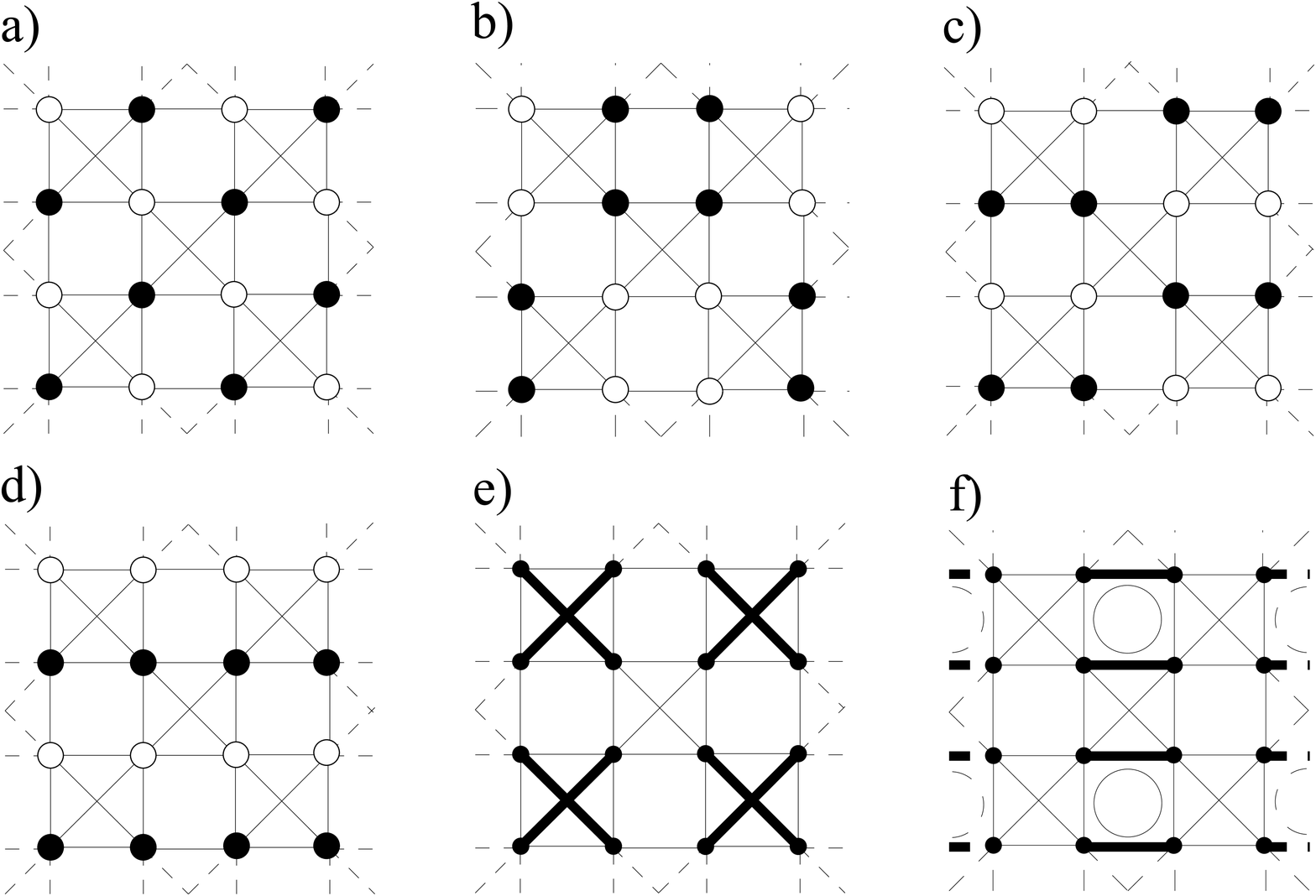}}
\caption{ Various ordered phases on the checkerboard lattice explored in this work; 
N\'{e}el order with a) ${\bf Q}=(\pi,\pi)$, b) ${\bf Q}=(\pi/2,\pi/2)$ (N\'{e}el$^*$), 
c) ${\bf Q}=(\pi/2,\pi)$, d) ${\bf Q}=(0,\pi)$ (stripe). Open (solid) circles denote 
down-spins (up-spins); (e) crossed-dimer order where thick (thin) diagonal lines represent 
strong (weak) bonds; and (f) P-VBS phase with strong dimer-dimer correlation between 
parallel bonds of uncrossed plaquettes marked by big circles.}
\label{fig:orders}
\end{figure}

The situation in the limit of weakly coupled crossed chains ($J'\gg J$) is less clear. 
There are at least two proposals for the ground state in this region of the parameter 
space; the first is the 2D spin-liquid ground state (sliding Luttinger liquid) characterized 
by the absence of long-range order and by elementary excitations being massless deconfined 
spinons.~\cite{O_Starykh_02} This idea is supported by an exact diagonalization study of Sindzingre 
{\em et al.},\cite{P_Sindzingre_02} which suggests a range of $J/J'=[0-0.8]$ for the 
1D behavior. However, their calculations suffer from strong finite-size effects even 
with $36$ sites due to the quasi-1D nature of the problem. The second is the crossed-dimer (CD) 
phase suggested by Starykh {\em et al.}~\cite{O_Starykh_05} They argued that, in the CD 
phase, staggered dimer correlations, which have a power-law decay with distance in a 
perfect 1D system, are stabilized when a weak interchain interaction ($J$) is present. 
As depicted in Fig.~\ref{fig:orders}(e), in this phase, the ``strong'' (positive) dimers 
from perpendicular chains meet at the same crossed plaquette. This scenario is 
in agreement with the results of Arlego {\em et al.},~\cite{M_arlego_07} who examined 
this idea by means of series expansion in terms of $J$ and $J'$ connecting the blocks 
of crossed dimers. Including results from other works, Starykh {\em et al.}~\cite{O_Starykh_05} 
also mapped out the global zero-temperature phase diagram of the system with respect 
to the ratio of $J$ and $J'$ and discussed the possibility of
a magnetically ordered phase being present in the transition between the CD phase and 
the P-VBS phase. This so-called N\'{e}el$^*$ phase is the long-range ordered phase with 
diverging susceptibility at ${\bf Q}=(\pi/2,\pi/2)$ [see Fig.~\ref{fig:orders}(b)]. 
Most recently, using a two-leg ladder to construct the 2D lattice in a density matrix 
renormalization group study, and by measuring various spin-spin correlations, 
Moukouri~\cite{S_Moukouri_08} confirmed most of these predictions for the phase 
diagram except that the magnetically ordered phase in the proximity of the CD phase 
has a wave vector ${\bf Q}=(\pi/2,\pi)$ instead of the ${\bf Q}=(\pi/2,\pi/2)$ proposed in 
Ref.~\onlinecite{O_Starykh_05}. Sketches of the former order, along with the other 
orders explored here, are shown in Fig.~\ref{fig:orders}.

Most of the numerical calculations for the AFH model on the checkerboard lattice have
been done at zero temperature using finite clusters with periodic boundary condition.
As discussed above, some of the early works~\cite{S_Palmer_01,J_Fouet_03,P_Sindzingre_02} 
helped shape theories that describe  ground-state properties such as the P-VBS.
However, a systematic study of finite-temperature properties in the thermodynamic limit, 
more relevant to experiments, has been missing. Our goal in this study is to explore
the thermodynamic properties of this model and address the finite-temperature behavior 
of the susceptibilities to the ordered phases proposed for the ground state and described
above.

\begin{figure}[t]
\centerline {\includegraphics*[width=1.3in]{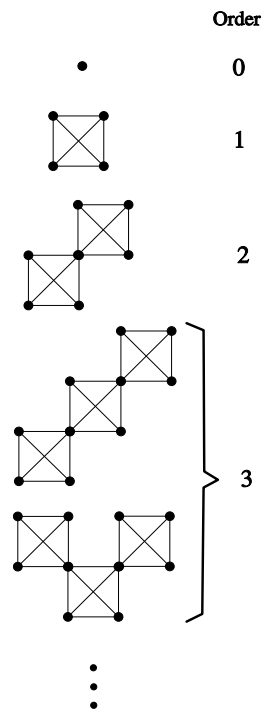}}
\caption{Clusters generated in the first four orders of NLCE with a square 
building block on the checkerboard lattice.}
\label{fig:SQ}
\end{figure}

We employ the numerical linked-cluster expansions (NLCEs),\cite{M_rigol_06,M_rigol_07a} 
along with exact diagonalization of finite clusters, to calculate thermodynamic 
properties of the system in different regions of the parameter space. We study the change 
in behavior of energy, entropy, specific heat, and several susceptibilities as $J$ and $J'$ 
vary. We find that the high-temperature peak in specific heat is strongly suppressed in the 
case of maximum classical frustration, $J'=J$. 
Consistently, we see large amounts of unquenched entropy in this region, signaling the 
possibility of a second peak in specific heat. Our study of different susceptibilities includes 
the staggered susceptibility, which for $J'/J\leq 0.75$ continues to grow as the temperature 
is lowered, suggesting that the ground state is N\'{e}el ordered with ${\bf Q}=(\pi,\pi)$ 
in this region. We also study the susceptibility to the the P-VBS phase using relevant order 
parameters and find that it is largest for $J'\sim J$. In the limit of weakly coupled 
crossed chains, and down to the lowest temperatures we can access, the dominant correlations 
belong to the N\'{e}el$^*$ and stripe phases.

The paper is organized as follows: In Sec.~\ref{sec:model}, we present the model and 
briefly discuss NLCEs, and the extrapolation techniques, along with the clusters utilized 
in the exact diagonalizations. The results are presented in Sec.~\ref{sec:results}, and a 
summary and conclusions are provided in Sec.~\ref{sec:summary}.

\section{model and numerical approach}
\label{sec:model}

\subsection*{The Hamiltonian}

The AFH Hamiltonian can be written as

\begin{equation}
H=J\sum_{\left <i,j\right >}{\bf S}_i\cdot{\bf S}_j + 
J'\sum_{\left <\left <i,j\right >\right >}{\bf S}_i\cdot{\bf S}_j,
\end{equation}
where ${\bf S}_i$ is the spin-$1/2$ vector at site $i$, and $\left <i,j\right >$ 
denotes bonds between NN sites $i$ and $j$. $\left <\left <i,j\right >\right >$
denotes bonds between NNN sites $i$ and $j$ on every other square in a 
checkerboard pattern.

\subsection*{Numerical Linked-Cluster Expansions}

\begin{table}[!b]
\caption{Size and number of topologically distinct clusters up to the sixth order 
of the square expansion.}
\begin{tabular}{ccc}
\hline\hline
\ \ \ \ \ \ Order\ \ \ \ \ \ & \ \ \ \ Number\ of sites\ \ \ \ \ &\ \ Number\ of clusters \ \ \\
\hline
  0  &      1  &\qquad  	     1 \\
  1  &      4  &\qquad  	     1 \\
  2  &      7  &\qquad  	     1 \\
  3  &     10  &\qquad  	     2 \\
  4  &     12  &\qquad  	     1 \\
  4  &     13  &\qquad  	     4 \\
  5  &     15  &\qquad  	     1 \\
  5  &     16  &\qquad  	    10 \\
  6  &     17  &\qquad  	     1 \\
  6  &     18  &\qquad  	     7 \\
  6  &     19  &\qquad  	    23 \\
\hline\hline
\end{tabular}
\label{tb:sq}
\end{table}

NLCEs are linked-cluster expansion methods which allow one to calculate the partition 
function and other observables, per lattice site, in the thermodynamic limit at finite 
temperatures. The information for these quantities at a given temperature is built up by
calculating contributions from all the clusters, up to a certain size, that can be embedded 
in the infinite lattice. Unlike high-temperature expansions (HTEs), each cluster is solved 
exactly using full diagonalization algorithms. Hence, NLCEs have a region of convergence 
which extends beyond that of HTEs. Depending on the type of ordering that 
occurs in the system at low temperatures, NLCEs can remain converged down to surprisingly 
low temperatures. Examples of these can be seen in the case of geometrically frustrated magnetic 
systems such as the Kagom\'{e} lattice, where there is no long-range magnetic 
ordering.~\cite{M_rigol_06,M_rigol_07a,M_rigol_07c} As in other series expansion approaches, 
we use extrapolation techniques to perform the summation of existing orders to further 
decrease the temperature of convergence, often gaining access to regions where most 
of the interesting phenomena take place. More details about these extrapolations can be
found in the following subsection and references therein.

Depending on the symmetry of the lattice and properties of the model, the generation of 
clusters in NLCEs can be done using different building blocks. These include the usual 
bond expansion, site expansion, triangle or square expansions, etc.\cite{M_rigol_07a} In this paper, 
we focus on the square expansion, which offers a particularly convenient approach in constructing 
the checkerboard lattice, i.e., by tiling it with crossed squares. In this picture, the first order 
in the expansion has a single crossed square, the second order has two crossed squares, 
and so on. The first four orders, including the zeroth order with a single site, are shown
in Fig.~\ref{fig:SQ}.

\begin{figure}[t]
\centerline {\includegraphics*[width=3.3in]{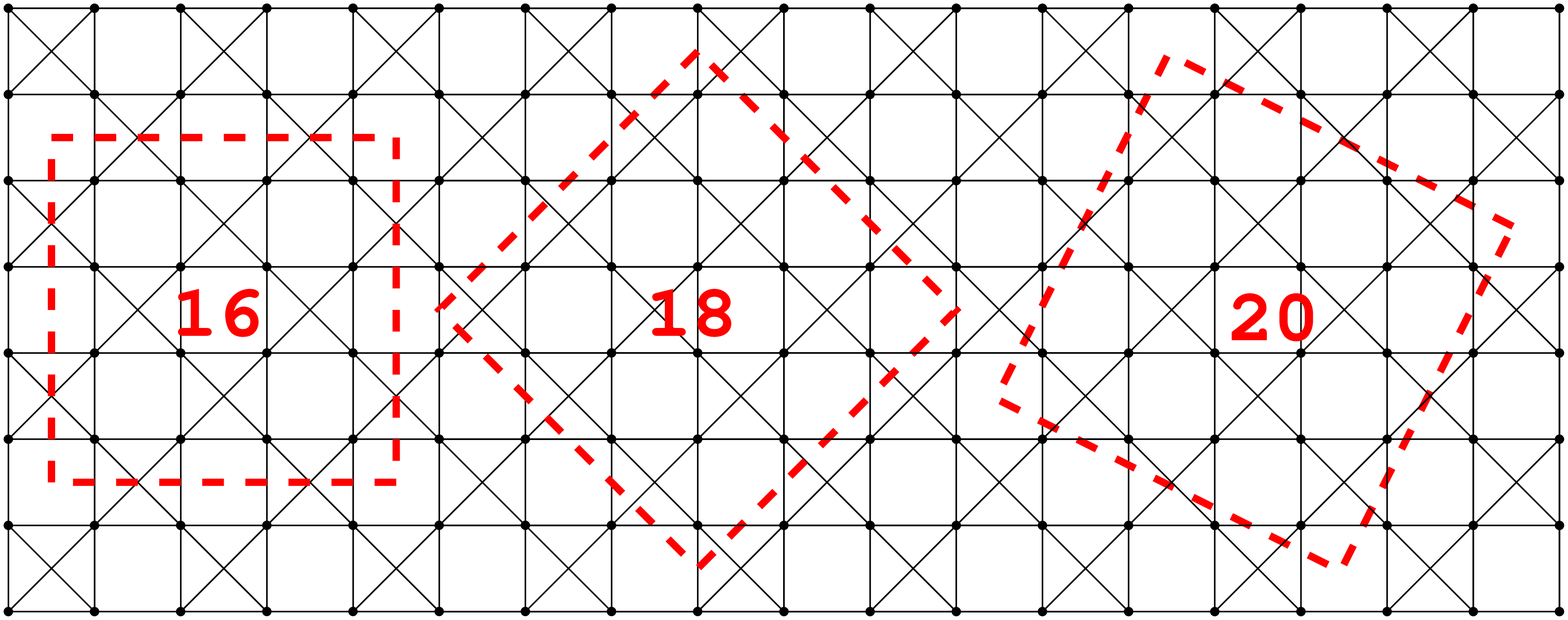}}
\caption{(Color online) Periodic clusters on the checkerboard lattice used in our 
finite-size exact-diagonalization calculations. The number inside each cluster represents 
its size.}
\label{fig:CB}
\end{figure}

In the square expansion, the maximum number of sites of a cluster in the $n{\text{th}}$ 
order is $3n+1$. Also, the number of topologically distinct clusters increases 
exponentially as the order increases. The number of clusters of each size, which need 
to be considered up to sixth order, is shown in Table~\ref{tb:sq}. Note that, out of $31$ 
clusters in the sixth order, $23$ have $19$ sites, $7$ have $18$ sites, and $1$ has $17$ 
sites. Since the clusters have open boundaries, no translational symmetries can be used 
to block-diagonalize the Hamiltonian matrix. This restricts the calculations to sixth 
or fewer orders, where, by using the conservation of the total spin in the $z$ direction, 
we have to diagonalize matrices with linear size as large as ${19 \choose 9}=92,378$. 
This is nearly impossible using serial LAPACK subroutines on single-processor 
machines given memory restrictions and the time needed for such huge diagonalizations. 
Therefore, most of the calculations have been performed on parallel computers using 
SCALAPACK routines.

Where possible, we compare results from NLCEs to those from exact diagonalization of 
finite clusters with periodic boundary conditions (ED) to build intuition about the 
finite-size effects that might have influenced results of previous studies. These clusters, 
with $16$, $18$, and $20$ sites, are shown in Fig.~\ref{fig:CB}. We use translational 
symmetries that are allowed on the checkerboard lattice and are not prohibited by the 
symmetries of the order parameters in the broken symmetry cases. The largest matrix 
we had to diagonalize in this case was for the $20$-site cluster, which had a linear 
dimension of $36,956$.

\subsection*{Extrapolations}

\begin{figure}[t]
\centerline {\includegraphics*[width=3.3in]{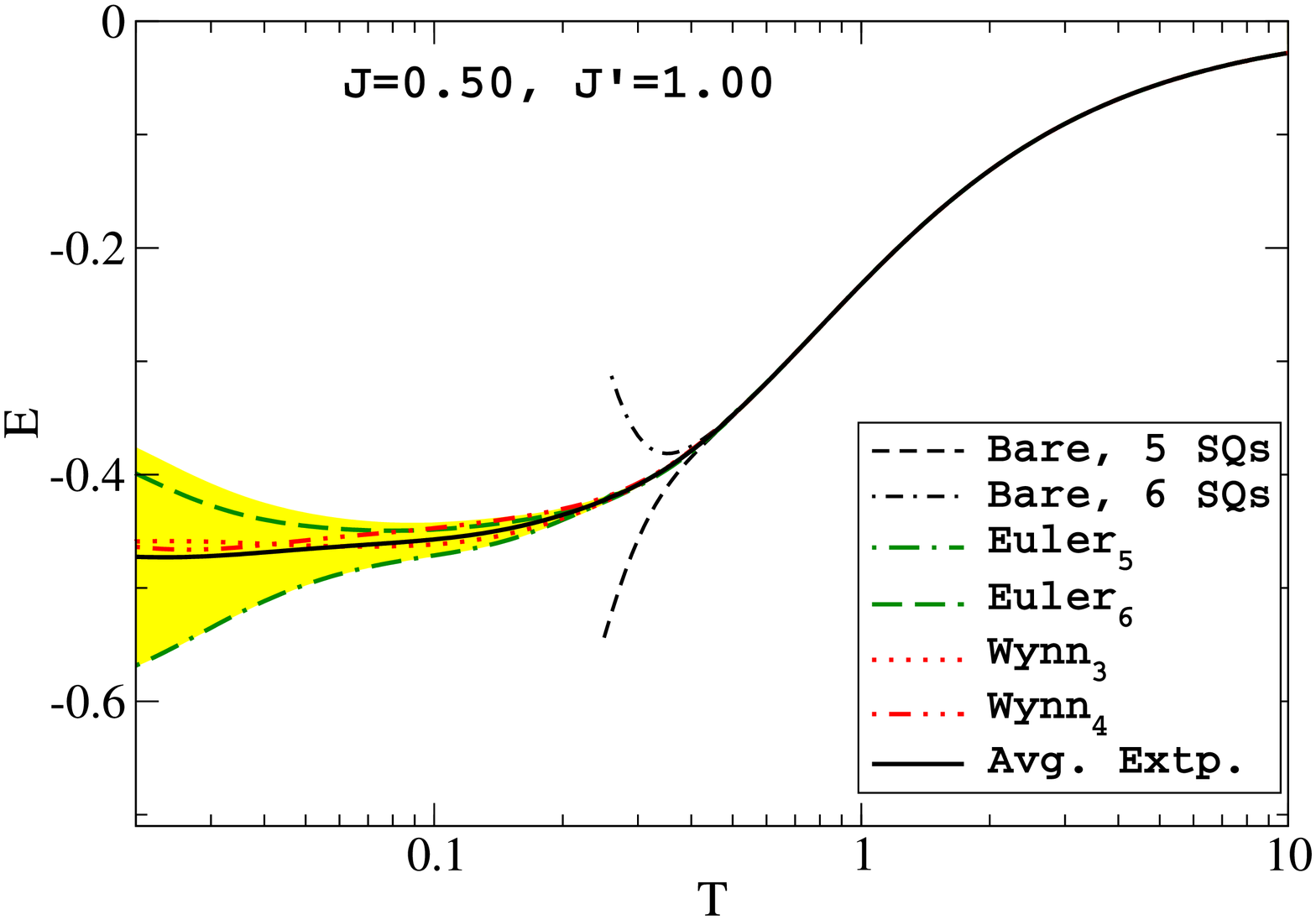}} 
\caption{(Color online) Energy per site vs temperature for the AFH  model on the 
checkerboard lattice with NN and NNN exchange interactions $J=0.50$ and $J'=1.00$, 
respectively. The thin dashed and dot-dashed 
lines represent the bare NLCE sums up to the fifth and sixth orders of the square expansion.
The solid line shows the average of the last two terms in the Euler and Wynn extrapolations with 
the shaded (yellow) area representing the ``confidence limit" where all the extrapolations 
lie. The unit of energy is $J'$.}
\label{fig:extrapolation}
\end{figure}

Measurements from all the clusters of every NLCE order are grouped together before summing 
different orders either regularly (bare sums) or by using Euler~\cite{Euler} or Wynn~\cite{Wynn} 
sequence extrapolation algorithms. (For a detailed description of these algorithms
see Ref.~\onlinecite{M_rigol_07a}.) In the Euler sum, one can choose to have bare 
sums up to a particular order before using the Euler algorithm for the remaining
orders. Here, we apply the Euler sum to the last four, three, two, and one terms. We find 
that the one with three Euler sums is generally the best (more physically sensible). In 
the Wynn sum, we can have one or two cycles of improvement, each eliminating two terms, 
leaving us with four and two terms, respectively, out of the initial six. Because of the 
small number of terms in the Wynn sum, we find that using only one cycle yields a more 
reliable outcome. Hence, unless otherwise mentioned, we show results throughout this 
paper for the Wynn sum with one cycle and Euler sum for three terms.

The behavior of these extrapolations can be seen in Fig.~\ref{fig:extrapolation}, where 
we show, as an example, the energy per site ($E$) versus temperature for $J=0.50$ and 
$J'=1.00$. We also include the bare sums up to the fifth and sixth orders, which start 
diverging around $T=0.4J'$. As expected, the results from the Euler and Wynn sums show a less 
divergent behavior and extend the region of convergence to lower temperatures. To have 
a rough estimate for energy at temperatures not accessible by bare NLCE sums, we take 
the average of the last two terms in the Euler and Wynn sums (solid line). All these 
four extrapolations lie in the shaded (yellow) region which can serve as the ``confidence 
limit.'' We refer to this region around the average as the error bar, although it by no means 
represents statistical error bars. Below the temperature where bare NLCE sums diverge, the 
extrapolations' average is not guaranteed to be the exact result in the thermodynamic limit. 
However, along with the error bars, it serves as an estimate of the desired quantity.

\section{results and discussions}
\label{sec:results}

Here we study thermodynamic properties such as total energy, entropy, specific heat, and
several magnetic susceptibilities for a range of parameters, sweeping different
regions of the phase diagram, from the simple square lattice without the NNN interaction
to near the 1D limit where NNN interactions dominate. For most of these
quantities, we show results for $J'=0.00$, $0.25$, $0.50$, $0.75$, and $1.00$ when $J=1.00$ 
and $J=0.75$, $0.50$, and $0.25$ when $J'=1.00$. The unit of energy is set to max($J,J'$).

\subsection*{Energy, Entropy, Specific Heat and Bulk Susceptibility}

\begin{figure}[t]
\centerline {\includegraphics*[width=3.3in]{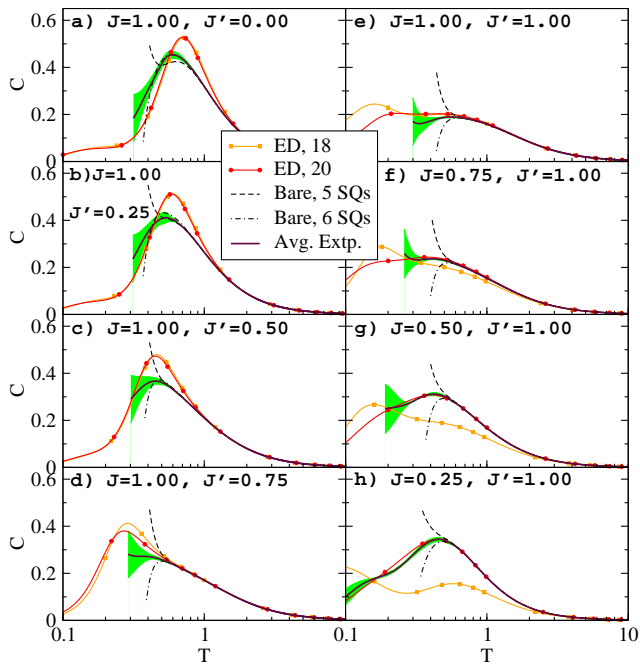}} 
\caption{(Color online) Specific heat vs temperature for various 
$J$ and $J'$: (a-d) $J'<J$ and (e-h) $J'\geq J$. For 
comparison, results from ED with $18$ and $20$ sites are shown. The first peak is 
captured for all cases after extrapolation. The NLCE results are cut off roughly 
where the error bars exceed $0.1$. The unit of energy is set to max$(J,J')$.}
\label{fig:SH}
\end{figure}

The specific heat per site ($C$) provides valuable information about the state of the system 
in different regions of the parameter space. In Fig.~\ref{fig:SH}, we show this quantity
after extrapolations of NLCE results for a range of values of $J'/J$. For comparison, results 
from ED with $18$ and $20$ sites are also shown. The highest peak appears for the simple 
square lattice with no frustration [see Fig.~\ref{fig:SH}(a)]. One can see that the bare NLCE 
results for fifth and sixth orders (dashed and dot-dashed lines, respectively) start deviating 
at $T\sim 0.8$, where the antiferromagnetic correlations presumably exceed the linear 
size of our biggest clusters. However, the average extrapolation captures a peak around $T=0.6$. 
More interestingly, both ED curves depart from the exact curve at a temperature greater than 
$J$ and show almost no improvement by increasing the cluster size from $16$ to $20$, with a 
position of the peak which is at slightly higher temperature. (The 16-site results are not shown.)

As $J'/J$ increases to $0.5$ [Figs.~\ref{fig:SH}(b) and (c)], the peak in specific 
heat broadens, its maximum value decreases, and the temperature at which the 
latter is reached also decreases. Due to the increase in frustration,
the AF correlations are suppressed and ED more accurately predicts the location 
of the peak while still overestimating its value. For the same reason, the convergence
in the bare NLCE sums is extended from $T\sim 0.8$ for $J'=0.0$ to $T\sim 0.5$ for 
$J'=0.5$. Further increasing $J'$ to $0.75$ [Fig.~\ref{fig:SH}(d)] changes
these features qualitatively by strongly suppressing the peak. In ED, the peak is 
pushed to lower temperatures ($T\sim 0.3$) and the agreement with exact NLCE results
can be seen down to lower temperature ($T\sim 0.5$) where the bare NLCE sums 
also diverge. These observations are consistent with results from previous studies 
that find a transition at zero temperature from the magnetically ordered N\'{e}el phase 
to a disordered phase for $J'\gtrsim 0.75J$.~\cite{r_singh_98,b_canals_02}

\begin{figure}[t]
\centerline {\includegraphics*[width=3.3in]{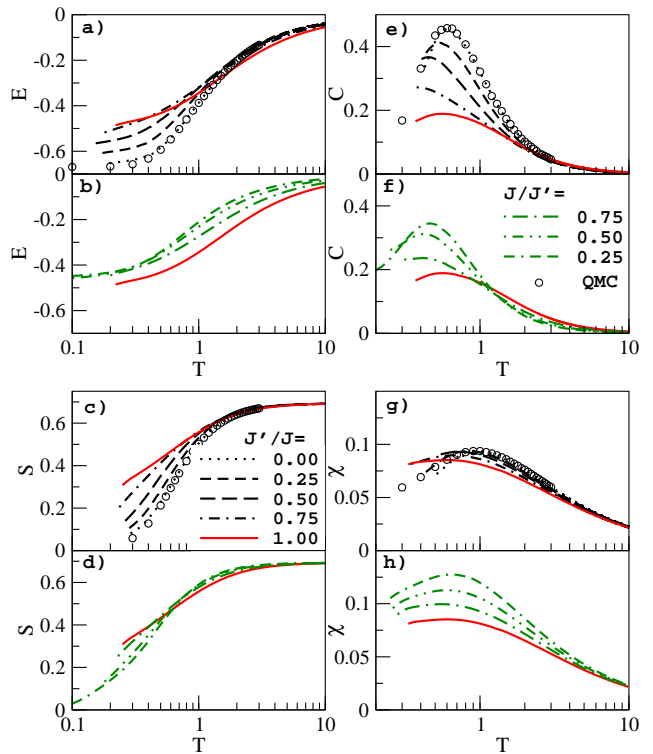}} 
\caption{(Color online) The evolution of (a, b) energy, (c, d) entropy, (e, f) specific heat, 
and (g, h) bulk susceptibility per site as $J$ and $J'$ change. These results are taken 
from the average extrapolations of NLCE and are cut off where the error bars reach 
$10\%$ or less. Circles in (a), (e), and (g) are the data from a large-scale quantum Monte Carlo (QMC) 
simulation for $J'=0$ (Ref.~\onlinecite{wessel}). Circles in (c) are the result 
of a direct integration of $C/T$ over temperature using the QMC results in (e), plus an 
additive constant to recover the infinite-temperature entropy, i.e., to account for 
the missing low-temperature tail of the specific heat.
The statistical error bars for the QMC are smaller than the symbols and are not shown.}
\label{fig:All}
\end{figure}

As expected, the minimum peak value is seen for the fully frustrated case 
of Fig.~\ref{fig:SH}(e) where $J'=J$ [see also Figs.~\ref{fig:All}(e) and \ref{fig:All}(f)]. Although 
ED is in good agreement with NLCEs for $T>0.5$, one can see significant finite-size
effects at lower temperatures between the $18$- and $20$-site clusters. The 
integral of $C/T$ for the temperature range shown for the average
extrapolation curve only recovers about half of the entropy at infinite temperature,
whereas $88\%$ is recovered for the case of Fig.~\ref{fig:SH}(a) with no frustration.
At $T=0.3$, the specific heat shows the tendency to develop a second peak. This 
tendency can be seen in both the NLCE and the ED results and, along with the fact 
that there is a huge amount of unquenched entropy already at $T\sim 0.3$ [see 
Figs.~\ref{fig:All}(c) and \ref{fig:All}(d)], strongly suggests that there is a second peak in specific 
heat at $T<0.3$.

As the value of $J/J'$ decreases from $1$, the peak in specific heat, shown in 
Figs.~\ref{fig:SH}(f)-\ref{fig:SH}(h), increases again and the 2D system starts to 
behave more and more like a 1D chain. This can be inferred from the dramatic 
finite-size effects in ED. While the $20$-site cluster can recover the NLCE results
with relatively good accuracy, the results for the $18$-site cluster start deviating 
from NLCE at temperatures as high as $2.0$. This can be understood from the fact 
that in the limit of decoupled crossed chains, $J=0$, the $18$-site cluster contains
six decoupled periodic chains, each consisting of only three sites, whereas the $20$-site 
cluster, contains two $10$-site decoupled chains. Note that not only are the 1D decoupled 
chains in the $18$-site cluster significantly smaller, but they also contain an odd number 
of chain sites, i.e., AF correlations are geometrically frustrated, and this strongly 
affects the results. Due to the quantum fluctuations, any long-range order is suppressed 
near the 1D limit, and so the extrapolations capture the specific heat with much smaller 
error bars for $J'/J=4$ as seen in Fig.~\ref{fig:SH} (h).

\begin{figure}[t]
\centerline {\includegraphics*[width=3.3in]{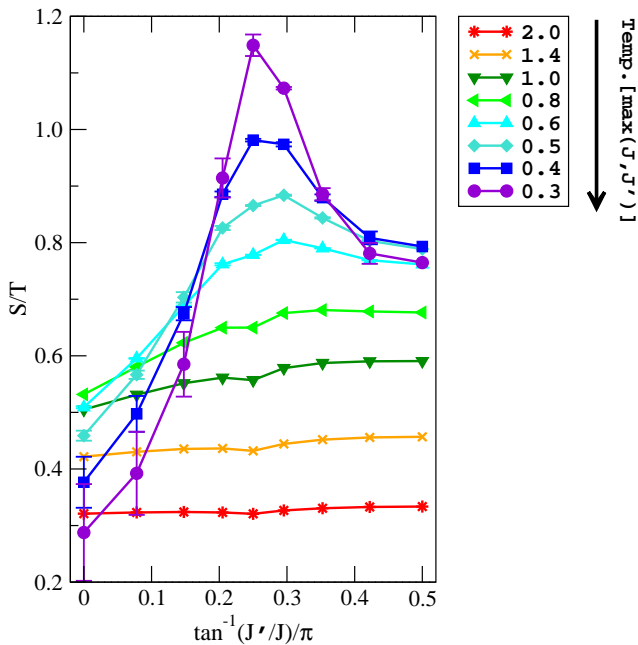}} 
\caption{(Color online) Entropy divided by temperature as a function of frustration 
angle, $\text{tan}^{-1}(J'/J)$. A peak in entropy develops at $J=J'$ as temperature is 
lowered below $0.5$. The values and the error bars are taken from the average 
extrapolation of NLCE results.}
\label{fig:EN-over-T}
\end{figure}

In Fig.~\ref{fig:All}, we show the evolution of energy, entropy ($S$), specific heat, 
and uniform susceptibility ($\chi$) per site as the value of $J'/J$ changes. One can see 
that the energy per site at temperatures below $0.2$ increases monotonically as $J'$ 
increases and, as expected from the results in Fig.~\ref{fig:SH}, the low-temperature 
entropy [Figs.~\ref{fig:All}(c) and \ref{fig:All}(d)] is maximal in the case of $J'=J$. 
The previously discussed decrease of the maximum value of the specific heat by increasing 
$J'$ to $J$, followed by an increase for larger values of $J'>J$, is more clearly seen in 
Figs.~\ref{fig:All}(e) and \ref{fig:All}(f). Finally, Figs.~\ref{fig:All}(g) and \ref{fig:All}(h) 
show that the uniform susceptibility remains small in all regions with a downturn below 
$T=1.0$. We have also included results from a large-scale stochastic series expansion QMC 
simulation (circles) with up to $256\times 256$ spins for the unfrustrated case of 
$J'=0$~\cite{wessel} using directed loop updates.~\cite{a_sandvik_99,f_alet_05}
This is the only case that we consider where the low-temperature QMC calculation is not 
limited by the sign problem.

To better compare the behavior of the entropy in different regions, in 
Fig.~\ref{fig:EN-over-T}, we show the entropy divided by temperature as a function of the
frustration angle defined as $\phi=\text{tan}^{-1}(J'/J)$. By lowering the temperature 
below $~0.5$, the entropy develops a peak at $J'=J$ which persists down to the lowest 
accessible temperature (with reasonable error bars for all angles). In the square lattice 
limit, $\phi=0$, the specific heat, and therefore the entropy, are known to be quadratic 
in $T$ at low temperatures. As can be seen in this figure, our results are consistent 
with this finding for $T\leq 0.5$. However, by increasing $J'/J$ to $1.0$, this behavior changes 
completely and entropy decreases even more slowly than $T$. On the other hand, close to the 
1D limit, $\phi>0.4\pi$, the entropy has a linear region around $T=0.5$ below which it 
decreases faster than $T$, similar to the weakly-frustrated regions with small $\phi$.


\subsection*{Order Parameter Susceptibilities}

Other than the uniform susceptibility, which can be measured directly from the 
fluctuations of the total spin in the $z$ direction, other susceptibilities
per site are calculated using their definition as the second derivative of the free 
energy with respect to the field that couples to the corresponding order parameter 
($\mathcal{O}$):
\begin{equation}
\chi^{\mathcal{O}}= \frac{T}{N}\frac{\partial^2\text{ln}Z}{\partial h^2}\Bigl |_{h=0},
\end{equation}
where $N$ is the number of sites, $Z$ is the partition function, and $h$ is the field 
that couples to the order parameter in the new Hamiltonian,
$\hat{H'}=\hat{H}-h\hat{\mathcal{O}}$.
For example, we consider the following order parameter for N\'{e}el orderings with 
different wave vectors: 
\begin{equation}
\hat{\mathcal{O}}_{\text{N\'{e}el}}= \sum_{\bf R} e^{i {\bf Q.R}} S^z({\bf R}),
\label{eq:stg}
\end{equation}
where ${\bf Q}=(q_x,q_y)$, ${\bf R}$ runs over a Bravais lattice with the basis 
${\bf a}=(\frac{\pi}{q_x},0)$ and ${\bf b}=(0,\frac{\pi}{q_y})$, and $S^z({\bf R})$ 
is the total spin in the $z$ direction in the corresponding unit cell.

\begin{figure}[t]
\centerline {\includegraphics*[width=3.3in]{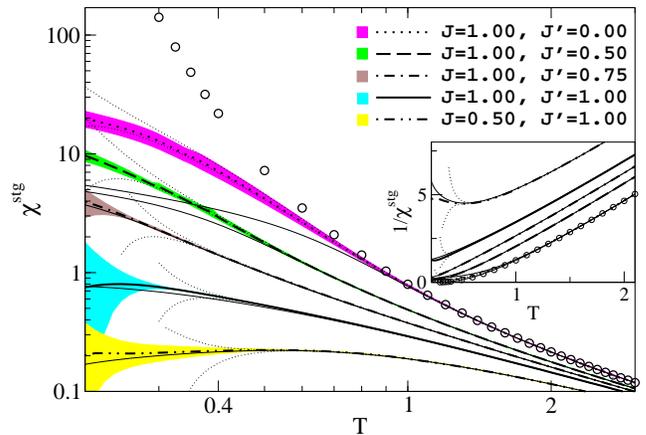}} 
\caption{(Color online) Log-log plot of extrapolated NLCE results for staggered 
susceptibility vs temperature.
By introducing $J'$,  the staggered susceptibility is suppressed. When $J=J'$, 
$\chi^{stg}$ is more than one order of magnitude smaller than in the case of $J'=0$. 
In the Euler extrapolation, only the last two terms have been used. The thin 
dotted lines are bare NLCE sums up to fifth and sixth orders. The thin solid 
lines represent the results from ED with 20 sites, and circles represent large-scale
QMC results for $J'=0$ (Ref.~\onlinecite{wessel}). The statistical error bars 
for the QMC results are smaller than the symbols and are not shown.}
\label{fig:stg}
\end{figure}

We find that the staggered susceptibility, $\chi^{stg}$ [${\bf Q}=(\pi,\pi)$], at low 
temperatures changes character when $J'/J$ is increased from $0.75$ to $1.00$. As can 
be seen in Fig.~\ref{fig:stg}, $\chi^{stg}$ continues to grow by decreasing temperature 
in the weakly-frustrated region and as long as $J'/J\leq 0.75$, but it shows a downturn at 
low $T$ for $J'/J\geq 1.00$. This is more clearly seen in the inset of Fig.~\ref{fig:stg}, 
where we have plotted the inverse of the staggered susceptibility versus temperature,
and is consistent with previous findings~\cite{P_Sindzingre_02,J_Fouet_03,r_singh_98,b_canals_02} 
which suggest that, in the latter region, the system no longer exhibits long-range N\'{e}el 
order. Note that the calculation of the staggered susceptibility 
for the unfrustrated case of $J'=0$ is one of the worst-case scenarios for NLCEs because the 
antiferromagnetic correlation length grows exponentially by decreasing the temperature. 
This can be realized by comparing the NLCE curve to the finite-size-converged (thermodynamic 
limit) QMC results (circles). Similar to the specific heat, the NLCEs results start deviating 
from the exact solution around $T=0.8$. Nevertheless, NLCE provides a far better estimate 
for this quantity at low temperatures than ED.

\begin{figure}[t]
\centerline {\includegraphics*[width=3.3in]{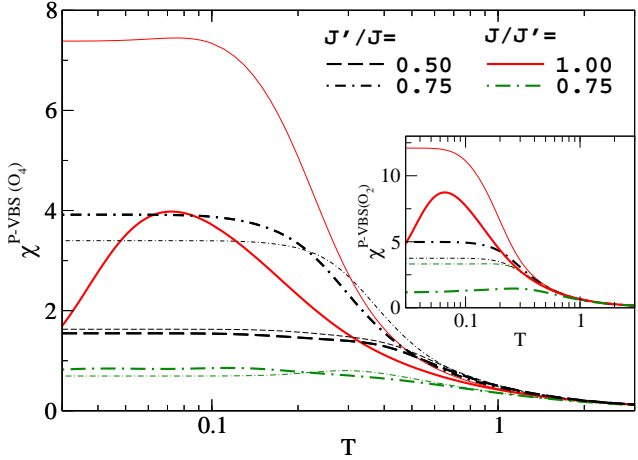}} 
\caption{(Color online) ED results for the susceptibility to the P-VBS order [see Eq.~(\ref{eq:DD})]
per site vs temperature. Thick (thin) lines are results for the $20$-site ($16$-site) 
cluster. The order is depicted in Fig.~\ref{fig:orders} (f). The inset shows the susceptibility to
the two-spin version of the plaquette order parameter as presented in Eq.~(\ref{eq:2}).}
\label{fig:ddm}
\end{figure}

\begin{figure}[t]
\centerline {\includegraphics*[width=3.3in]{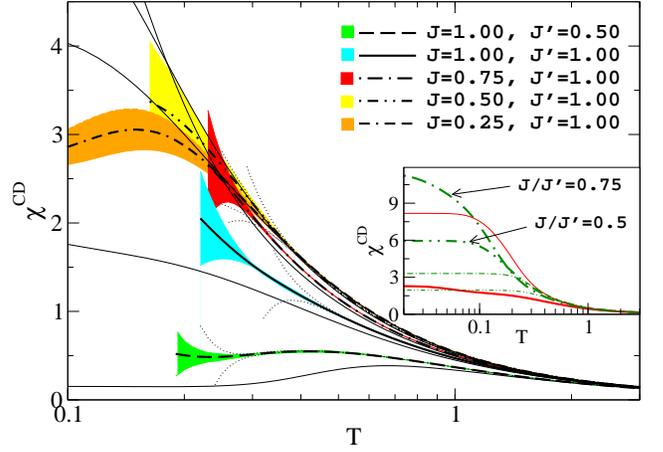}} 
\caption{(Color online) Susceptibility to the crossed-dimer order [see Eq.~(\ref{eq:CD)} 
and Fig.~\ref{fig:orders}(e)] 
per site vs temperature. Thin dotted lines are the last two orders 
of bare NLCE sums, and thin solid lines are the ED results with $20$ sites. 
In the Euler extrapolation, only the last two terms have been used. In the inset, 
lines are as in Fig~\ref{fig:ddm} with thick (thin) lines representing ED results 
for the $20$-site ($16$-site) cluster.}
\label{fig:cdm}
\end{figure}

\begin{figure}[t]
\centerline {\includegraphics*[width=3.3in]{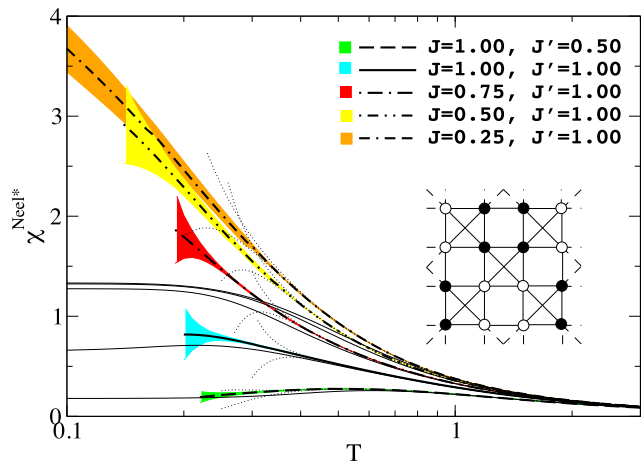}} 
\caption{(Color online) N\'{e}el$^*$ susceptibility [${\bf Q}=(\pi/2,\pi/2)$] per site 
vs temperature for a range of ratios of $J$ and $J'$.
The inset shows the corresponding magnetic order where the open (solid) circles
represent down-spins (up-spins) spins. Thin dotted lines are the last two orders 
of bare NLCE sums, and thin solid lines are the ED results with $16$ sites.}
\label{fig:nel}
\end{figure}

\begin{figure}[t]
\centerline {\includegraphics*[width=3.3in]{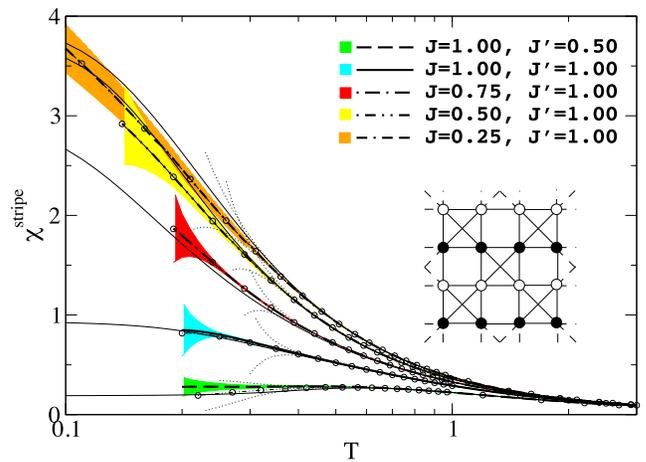}} 
\caption{(Color online) Susceptibility to the stripe order [${\bf Q}=(0,\pi)$] per site
vs temperature. The inset shows the corresponding magnetic order where the 
open (solid) circles represent down-spins (up-spins) spins. Thin dotted lines are the last two orders 
of bare NLCE sums, and thin solid lines are the ED results with $20$ sites. Circles are
NLCE results for the N\'{e}el$^*$ order (Fig.~\ref{fig:nel}).}
\label{fig:nl4}
\end{figure}

According to the Mermin-Wagner theorem,\cite{M-W} the Heisenberg model with finite-range 
exchange interactions in two dimensions cannot undergo a phase transition to a long-range 
ordered state at finite temperature by breaking a continuous symmetry. However, in light 
of the recent analytical and numerical 
predictions for the ground-state phases of this system, we calculate the finite-temperature 
susceptibilities associated with various order parameters to study their behavior as the 
temperature is lowered. These susceptibilities are shown in Figs.~\ref{fig:ddm} to \ref{fig:nl4} 
in a low-temperature window for the relevant values of $J$ and $J'$.

In Fig.~\ref{fig:ddm}, we show the susceptibility to a plaquette order which is expected
to be large in the P-VBS phase around $J'=J$. Fouet {\em et al.}~\cite{J_Fouet_03} argued 
that the ground-state wave function in this phase is the symmetric combination of the pairs 
of singlets on parallel bonds of the uncrossed plaquettes. Based on that, we consider the 
following four-spin order parameter:
\begin{equation}
\hat{\mathcal{O}}_{4}= 32\times \sum_{\bigcirc} {\bf S}_{l1}\cdot{\bf S}_{l2} \ 
{\bf S}_{l3}\cdot{\bf S}_{l4} ,
\label{eq:DD}
\end{equation}
where $l$ is the position of every other uncrossed square, marked by a circle in Fig.~\ref{fig:orders} (f). 
The spin numbers around each of these squares are such that $1$ and $2$ (and 
therefore, $3$ and $4$) are nearest neighbors. Since this kind of order involves uncrossed squares, 
NLCEs in crossed squares are not suited to measure the corresponding susceptibility, 
and so we have obtained results only from ED. They show that this susceptibility is largest in the 
region around the maximum classical frustration. However, significant finite-size effects are seen, 
especially for the $J'=J$ case. In this region, the results for the $16$-site cluster deviate 
from those for the $20$-site cluster when $T<1.0$, with the susceptibility being roughly a factor 
of $2$ larger at $T\sim 0.1$ for the $16$-site cluster. Interestingly, for the $20$-site 
cluster, the susceptibility shows a significant decrease by further decreasing temperature below
$T=0.07J$. Note that most of the thermodynamic quantities, such as the specific 
heat and other susceptibilities calculated using ED (even with $20$ sites), deviate from their 
exact NLCE counterparts (bare sums) starting from temperatures as high as $0.5$ in this 
parameter region. So, the peak feature is expected to be a consequence of the finite-size 
nature of the calculations. We tested a more sophisticated order parameter suggested
in Ref.~\onlinecite{J_Fouet_03} to better capture the P-VBS phase, namely, the four-spin 
cyclic permutation operator ($P_4+P_4^{-1}$),\cite{D_Thouless_65} and found the same qualitative 
results as for $\hat{\mathcal{O}}_{4}$ after rescaling.

Alternatively, one can define a simple two-spin order parameter as the sum of strong NN 
bonds around every other empty plaquette and weak NN bonds elsewhere to describe this phase:
\begin{equation}
\hat{\mathcal{O}}_{2}=\sum_{\Box}(-1)^{l_x} ({\bf S}_{l1}\cdot{\bf S}_{l2}
+{\bf S}_{l2}\cdot{\bf S}_{l3}+{\bf S}_{l3}\cdot{\bf S}_{l4}+{\bf S}_{l4}\cdot{\bf S}_{l1}),
\label{eq:2}
\end{equation}
where $l$ is the position of each empty square ($\Box$) in units of the NN lattice 
spacing and we have numbered the spins in each square clockwise, starting from the bottom 
left corner. The resulting susceptibilities for three values of $J'/J$ around the 
fully frustrated region are plotted in the inset of Fig.~\ref{fig:ddm} and show similar 
trends as their four-spin counterparts.

By decreasing $J/J'$ to $0.5$, we find that the low-temperature susceptibility to the CD
order is enhanced (Fig.~\ref{fig:cdm}). To calculate this susceptibility, we consider 
the following order parameter:
\begin{equation}
\hat{\mathcal{O}}_{\text{CD}}= 2\times \sum_{\boxtimes}(-1)^{l_x} ({\bf S}_{l1}\cdot{\bf S}_{l3}
+{\bf S}_{l2}\cdot{\bf S}_{l4}),
\label{eq:CD}
\end{equation}
where $l$ is the position of each crossed square ($\boxtimes$) and spin numbering is the 
same as in Eq.~(\ref{eq:2}) so that, ${\bf S}_1$ and ${\bf S}_3$ (or ${\bf S}_2$ 
and ${\bf S}_4$) are at the two ends of diagonal bonds. Although this susceptibility is 
significantly larger for $J'/J>1$, the extrapolated values for $J'/J=4$ exhibit a downturn 
at finite temperature.
The results from ED with $20$ sites overestimate the NLCE results at low $T$ for $J'>J$. 
However, we see significant finite-size effects between the $16$- and $20$-site clusters, 
shown in the inset of Fig.~\ref{fig:cdm}. We have checked the susceptibility to a closely 
related order parameter in which there is one strong diagonal bond on every crossed plaquette, 
[specifically, Eq.~(\ref{eq:CD}) with a minus sign between the two terms] and found a 
behavior qualitatively similar the CD susceptibility but with smaller values (not shown). 
Since the CD phase was predicted to exist for  $J'\gg J$~\cite{O_Starykh_05}, an 
interesting question posed by these results is whether the peak feature will eventually 
disappear for smaller values of $J/J'$ and one would find a susceptibility that always
increases with decreasing temperature. In this scenario, the relevant temperature at 
which the CD phase becomes dominant is $\mathcal{O}(J^2/J')$~\cite{O_Starykh_05}, 
which is beyond the convergence region of our current NLCE calculations.

We find that for large values of $J'/J>2$ (weakly coupled crossed chains), there are 
two magnetic orderings that are dominant at the lowest temperatures we can study. 
They are (i) the so-called N\'{e}el$^*$ order and (ii) stripes along the horizontal 
(or vertical) directions (Figs.~\ref{fig:nel} and \ref{fig:nl4}). The corresponding 
order parameters are defined in Eq.~(\ref{eq:stg}) with ${\bf Q}=(\pi/2,\pi/2)$ and 
${\bf Q}=(0,\pi)$, and are depicted in Fig.~\ref{fig:orders}(b) and \ref{fig:orders}(d), 
respectively. The former has been proposed theoretically as the candidate for this 
region.~\cite{O_Starykh_05} An intriguing observation is that the values for these two 
susceptibilities are hardly distinguishable, especially when $J'>J$. To illustrate 
the latter, we plot the NLCE results for the N\'{e}el$^*$ susceptibility against the 
stripe susceptibility in Fig.~\ref{fig:nl4} (circles). One can see that the relative 
difference is negligible for all values of $J'/J$ shown.

These results are consistent with what one would expect at intermediate temperatures 
in the limit of $J'\gg J$, because both orders are compatible with the antiferromagnetic 
correlations that develop along the diagonal chains. We note that for the ED with the 
$16$-site cluster, using adjacency matrices, one can show that the modified 
Hamiltonians, $\hat{H}'$, are identical for the two order parameters. It would have been 
interesting to compare ED results for both orders with larger system sizes; however, 
given the unit cell size for each order (eight sites for the N\'{e}el$^*$ and four sites 
for the stripe) and our computational limitations with increasing system sizes, those 
results are only available for the stripe order and are shown in Fig.~\ref{fig:nl4}. Resolving 
which order becomes dominant at lower temperatures will require the study of larger 
cluster sizes, both in NLCEs and in ED. It is worth mentioning that, based on numerical 
calculations, the stripe order was suggested to be the one relevant to the ground 
state of the $J_1-J_2$ model when $J_2 \gtrsim 0.6J_1$.\cite{S_Moukouri_08}

\begin{figure}[t]
\centerline {\includegraphics*[width=3.3in]{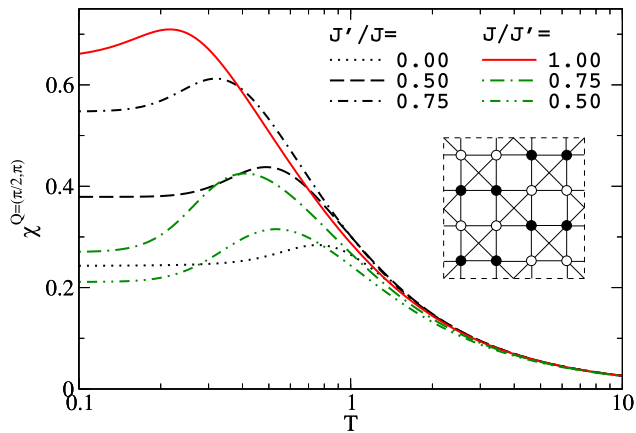}} 
\caption{(Color online) Susceptibility to the N\'{e}el order with ${\bf Q}=(\pi/2,\pi)$
per site, calculated using ED with $16$ sites, versus temperature. The inset shows 
the corresponding order where empty (full) circles represent down (up) spins.}
\label{fig:Neel3}
\end{figure}

We have explored another magnetic ordering suggested by Mokouri~\cite{S_Moukouri_08}
to be dominant in the limit $J'\gg J$. It has a wave vector of 
${\bf Q}=(\pi/2,\pi)$ and a unit cell of eight sites. Because of the breaking of certain 
symmetries of the lattice, the NLCE calculations for this case become much more 
expensive because one has to compute the physical properties of each cluster at different 
orientations and locations on the lattice to properly deal with the broken symmetry. 
(This applies to the previously mentioned orders as well, but, due to the presence of other 
symmetries, computations are less costly in those cases.) Thus, we only present results 
from ED with $16$ sites for this type of order. As shown in Fig.~\ref{fig:Neel3}, 
not only is this susceptibility smaller close to the 1D limit, 
but the maximum value, which belongs to the case of $J'=J$, is also much smaller than 
the maximum value seen for other orders with $16$ sites (see, e.g., Fig.~\ref{fig:nel}).
Therefore, a transition to this phase seems unlikely in any of the
parameter regions. The fact that this type of order is not favored close to the 1D limit is 
not surprising since, unlike in the N\'{e}el$^*$ or stripe ordered phases, spins on 
diagonal chains are not antiferromagnetically aligned.

\section{Summary}
\label{sec:summary}

We have calculated the thermodynamic properties of the AFH model on the checkerboard 
lattice using NLCEs and ED, and studied their behavior as the system crosses over
from a simple square lattice ($J'=0$) to the maximally frustrated planar 
pyrochlore lattice ($J'=J$) to the limit of one-dimensional crossed chains ($J'\gg J$).

We found that the peak value in the specific heat is suppressed as the frustration
increases (by increasing $J'/J$ from $0$ to $1$), with strong indications that
there is a second peak in the specific heat for $J'\sim J$. In the same region,  
finite-size effects in ED are minimal for temperatures above the convergence limit
of NLCE. In contrast, close to the 1D limit, ED results can vary significantly from 
one cluster to the other, depending on the size of periodic 1D chains that exist 
inside each 2D cluster. Consistent with the reduced specific heat, entropy is 
maximal when $J'=J$ at low temperatures with a decrease that is slower than $T$.

We calculated the susceptibilities to several magnetic and bond orderings to explore
the tendencies of the system toward different phases as the temperature is decreased.
By studying the staggered susceptibility, we found that the tendency toward N\'{e}el 
ordering with ${\bf Q}=(\pi,\pi)$ decreases appreciably when $J'/J\gtrsim 0.75$.
By increasing the NNN interaction, antiferromagnetic 
correlations along the diagonal chains become important and other types of order emerge.
To investigate this, we examined the susceptibility to the P-VBS order using ED, 
and found that it is largest for $J'\sim J$. We also
found large finite-size effects between $16$- and $20$-site clusters for $J'=J$.

We further explored the susceptibility of the CD order, which is larger for $J'>J$ but, 
according to the extrapolated NLCE results and for the values of $J$ and $J'$ considered
here, does not dominate at the intermediate temperatures accessible within our NLCEs. 
Finite-size effects between the $16$- and $20$-site clusters were found to be 
significant in the ED calculations for $J'\geq J$. When $J'>2J$, i.e., for weakly coupled 
crossed chains, we found fast increasing susceptibilities at intermediate temperatures 
to N\'{e}el$^*$ order with ${\bf Q}=(\pi/2,\pi/2)$, suggested by analytical results, 
and stripe order with ${\bf Q}=(0,\pi)$. Both of these orders are favored in this 
region due to the antiferromagnetically aligned spins along the chains.

\section{Acknowledgments}

This research was supported by the National Science Foundation (NSF) under Grant 
No.~OCI-0904597 and enabled by allocation of advanced computing resources, supported 
by the NSF. Part of the computations were performed on Ranger and Lonestar at the 
Texas Advanced Computing Center under Account No.~TG-DMR100026.
We thank Stefan Wessel for providing us with the QMC results and for helpful 
discussions. We are grateful to Rajiv R. P. Singh and Oleg Starykh for 
careful reading of the manuscript and their useful comments. 


\end{document}